\documentclass[aps,prl,reprint,groupedaddress]{revtex4-1} 
\usepackage{graphicx}
\usepackage{dcolumn} 
\usepackage{bm} 
\usepackage{color}

\begin{document}

\title{Polarization discontinuity driven two dimensional electron gas at
A$_2$Mo$_3$O$_8$/B$_2$Mo$_3$O$_8$ (A, B : Zn,Mg,Cd) interfaces}

\author{Tathagata Biswas} 
\affiliation{Center for Condensed Matter Theory, Department of Physics, Indian Institute of Science, Bangalore, 560012}  
\author{Manish Jain}
\email{mjain@iisc.ac.in} 
\affiliation{Center for Condensed Matter Theory, Department of Physics, Indian Institute of Science, Bangalore, 560012}

\date{\today}

\begin{abstract} 
We prospose a novel heterostructure system consisting of compounds with
chemical formula A$_2$Mo$_3$O$_8$ (A, B : Zn,Mg,Cd) that can host a two
dimensional electron/hole gas (2DEG/2DHG). We study spontaneous polarization
and piezoelectric properties of these compounds using first principles methods
and Berry phase approach. We show that these kind of heterostructures are very
stable due to extreamly low interfacial strain. The formation of a 2DEG/2DHG
has been investigated in case of Zn$_2$Mo$_3$O$_8$/Mg$_2$Mo$_3$O$_8$ and
polarization discontinuity has been found to be driving mechanism. The sheet
carrier densities and charge localization in these kind of heterostrcutures
have been found to be of the same order of magnitude in other well known system
that hosts 2DEG through similar mechanism, such as AlN/Al(Ga)N or ZnO/Zn(Mg)O. In addition 
to conventional applications of a 2DEG, these materials hold promise to
exciting pioneering technolgy such as piezo-phototronics using solar
radiation, as they are also capable of absorbing a significant fraction of it
due to low optical gap of $\sim$ 2 eV.   
\end{abstract}

%\pacs{Valid PACS appear here}

\maketitle

Transition metal oxides (TMOs) have stimulated a large amount of theoretical
and experimental research over the last few decades. These oxides
simultaneously possess spin, charge and orbital degrees of freedom originating
from their strongly-correlated open $d$-shell electrons. As a result, they
exhibit a variety of interesting properties such as Mott insulators, various
charge, spin and orbital orderings, metal-insulator transitions, multiferroics
and even superconductivity\cite{imada1998metal,rao1989transition}. The
interfaces in TMOs can offer even more versatile and unique emergent many-body
phenomenon. This is primarily due to broken spatial inversion symmetry and
enhanced electron correlation in two dimensions
\cite{chakhalian2012whither,chakhalian2014colloquium,hwang2012emergent}.  In
the past, experimental studies of TMO interfaces were hindered by difficulties
in growing defect-free single crystal of these materials as well as fabricating
clean interfaces at atomic scale. Recent advances in the angstrom-scale
layer-by-layer synthesis of multi-element compounds and improved expertise in
molecular beam epitaxy, metal-organic vapor deposition techniques has enabled
the exploration of a wide variety of TMO interfaces
\cite{schlom2008thin,chakhalian2014colloquium}. 

Two dimensional electron gas (2DEG) formation at TMO interfaces is an
interesting phenomenon that can serve as a testbed for understanding electron
correlations in low dimensions \cite{abrahams2001metallic}. Moreover, this phenomenon has promising
technological implications. Owing to it's unique transport properties, 2DEGs
can be used in power electronics, high mobility electron transistor (HEMT),
spintronics, optoelectronics and other future nano-electronics devices
\cite{hwang2012emergent,chakhalian2014colloquium}. There are three general
mechanisms that can create a two dimensional electron gas at oxide interfaces.
The first one involves a wide-band-gap/narrow-band-gap heterostructure  and
modulation doping\cite{anderson1960germanium,delagebeaudeuf1982metal}.
AlGaAs/GaAs interface is an example of this mechanism.  The second one is
driven by the polar catastrophe, which originates from the divergence of
electric potential.  This mechanism can be observed for example at the
LaAlO$_3$/SrTiO$_3$ interfaces
\cite{bristowe2014origin,ohtomo2004high,nakagawa2006some,thiel2006tunable,park2010creation}.
The third mechanism originates as a result of polarization discontinuity at the
interface of two materials having different spontaneous (or strain induced)
polarization. The two most studied examples of this mechanism are at the
Al$_{1-x}$Ga$_{x}$N/GaN and the Zn$_{1-x}$Mg$_{x}$O/ZnO interface. In this case
uncompensated bound charge at the interface creates an internal electric field
which confines any free carrier close to the interface resulting in a 2DEG
\cite{heikman2003polarization,tampo2008polarization,betancourt2013polarization}. 

In this work we propose a new heterostructure system of TMOs where a
polarization discontinuity driven 2DEG can be formed at the interface. The
group of TMOs we propose have a chemical formula A$_2$Mo$_3$O$_8$. This group
of materials have been synthesized using non-magnetic (Zn, Mg, Cd) as well as
magnetic (Fe, Ni, Co, Mn) \cite{mccarroll1957some} divalent cation, A. In this
study we focus on the materials that have non-magnetic cations --
Zn$_2$Mo$_3$O$_8$ (ZMO), Mg$_2$Mo$_3$O$_8$ (MMO) and Cd$_2$Mo$_3$O$_8$ (CMO).
We find that by changing the divalent cation
we can significantly change the piezoelectric properties as well as spontaneous
polarization of these materials. We show that a heterostructure which consists
of two of these compounds can form a 2DEG at there interface due to
polarization discontinuity.  We also show that the interfacial strain due to
lattice mismatch in this materials are extremely small, so one can expect to
make a very clean interfaces.  We use first-principles calculations based on
density functional theory (DFT) to calculate the piezoelectric constants and
spontaneous polarization of these materials as well as explore the
formation of of a 2DEG or two dimensional hole gas (2DHG) when they form a
heterostructure. We calculate the interfacial charge density and the electric
fields in the heterostructure and show that they are consistent with the
polarization discontinuity hypothesis. We find that the sheet carrier density
in these heterostructure systems to be similar to the conventional example
systems like Al$_{1-x}$Ga$_{x}$N/GaN or Zn$_{1-x}$Mg$_{x}$O/ZnO
\cite{heikman2003polarization,tampo2008polarization}.  Moreover,
A$_2$Mo$_3$O$_8$ class of compounds has been also studied in the literature
\cite{paranthaman1988photoelectrochemical,biswas2017optical} as a potential
photoabsorber.  The optical gaps of these materials ($\sim$ 2.0 eV) are
suitable for absorbing significant fraction of solar radiation. In principle,
one may be able to combine the piezoelectric properties of these materials with
their photo-electrochemical properties making them a potential choice for
exotic opto-electronics applications such as piezo-phototronics
\cite{wang2012progress} using solar radiation. 

\begin{figure}[h!]
\setlength{\unitlength}{0.1\textwidth}
\begin{picture}(4,4)
\put(0,2.1){\includegraphics[height=3cm] {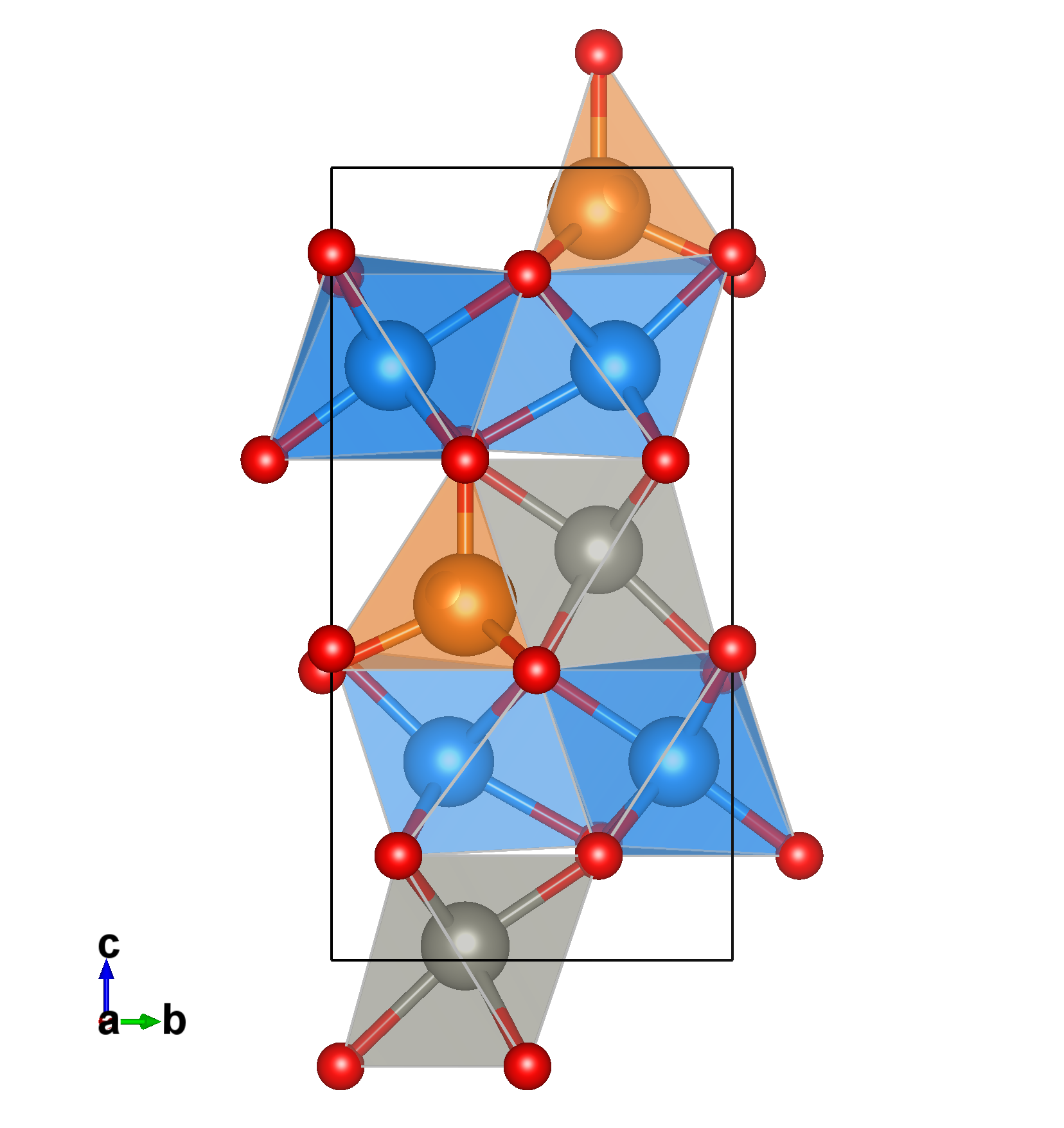}}
\put(2.1,2.1){\includegraphics[height=3cm] {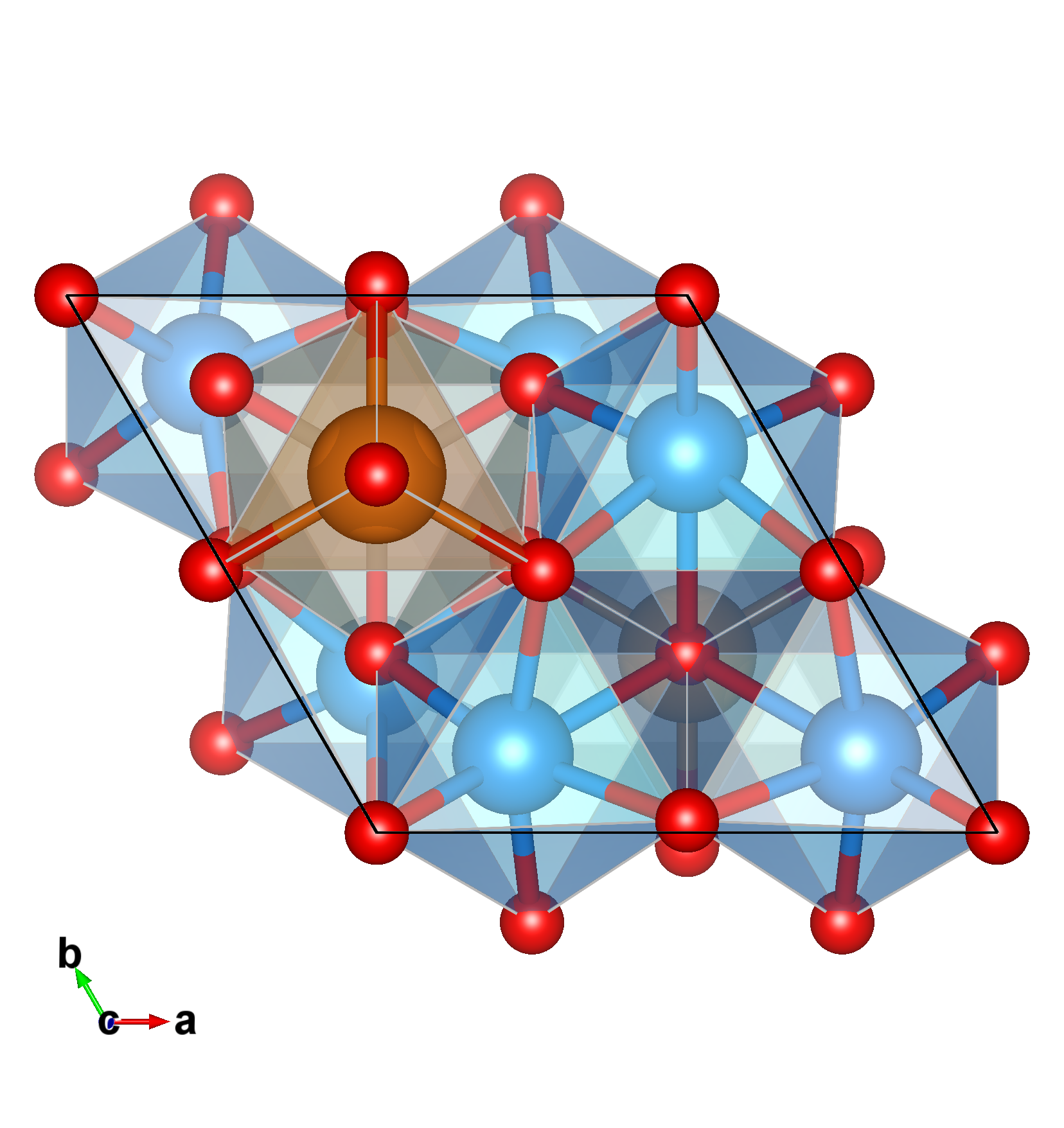}}
\put(-0.2,0.1){\includegraphics[height=3.2cm] {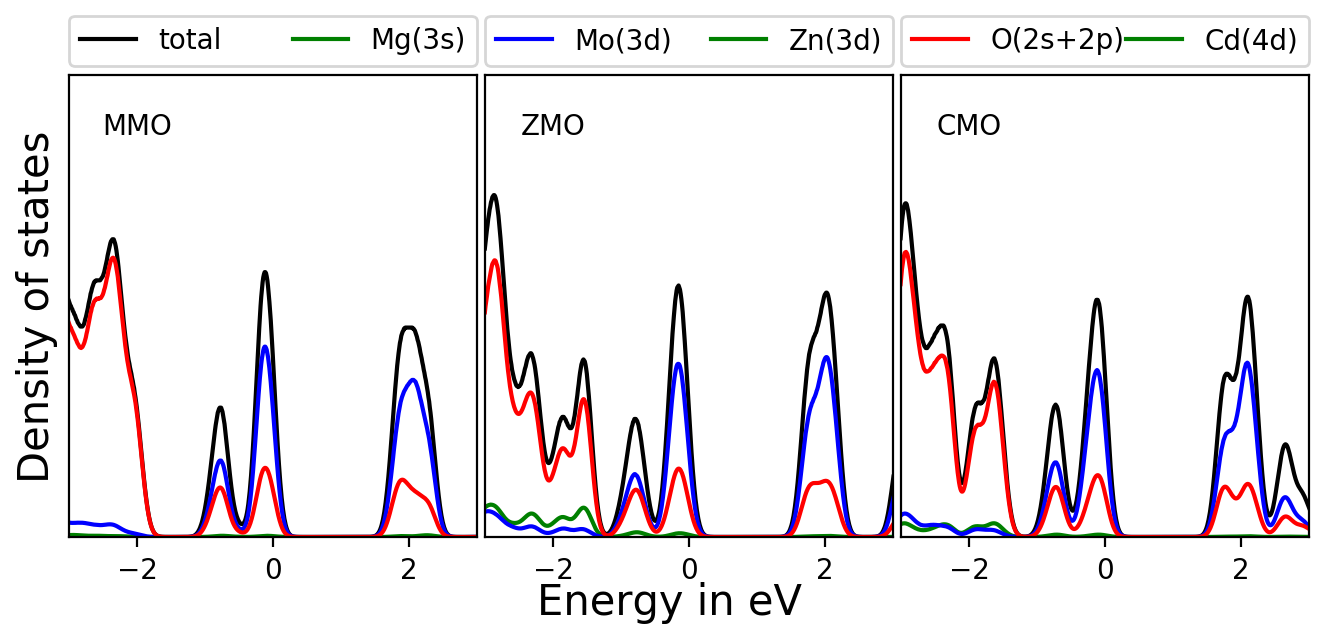}}
\put(0.7,2.0){(a)}
\put(3,2.0){(b)}
\put(2,0){(c)}
\end{picture}
\caption{ Crystal structure of A$_2$Mo$_3$O$_8$ compounds from (a)
[$2\bar{1}\bar{1}0$] (b) [$0001$] direction.  Yellow, grey, blue and red
spheres represents A$_{tetra}$, A$_{octa}$, Mo and O atoms respectively.
(c) Orbital resolved DOS of different A$_2$Mo$_3$O$_8$ compounds.
Valence band maxima has been set to zero in all cases.}
\label{fig1}
\end{figure}
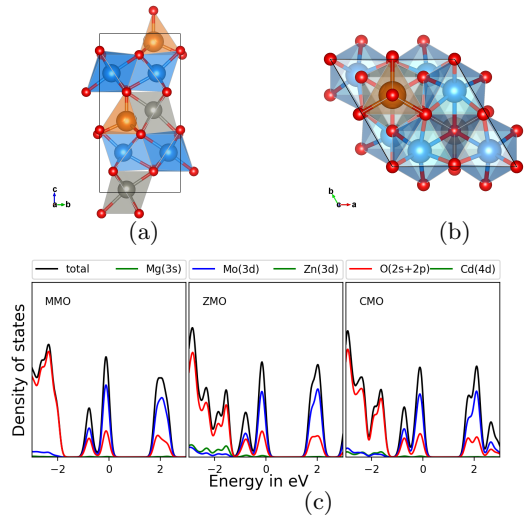

A$_2$Mo$_3$O$_8$ compounds crystallize in a hexagonal unit cell with space
group P6$_3$mc \cite{mccarroll1957some}. As this space group does not have
inversion symmetry, these materials can have a non-zero spontaneous
polarization.  Fig.~\ref{fig1}(a, b) shows the crystal structure of these materials
from two different directions. The crystal structure of these materials
consists of alternate layers of divalent cation (A) and Mo. A occupies both
tetrahedral (A$_{tetra}$) and octahedral (A$_{octa}$) sites whereas Mo occupies only octahedral sites. The
O atoms form layers between each A and Mo layers in a distorted hexagonal
closed pack structure. The stacking of O atoms in [$0001$] direction is in
$abac$ sequence.  These materials have been also categorized as metal oxide
cluster compounds as the three nearest in-plane Mo atoms form a strong bonds
between them and make a cluster. The existence of this strong bonds is
manifested as a smaller Mo-Mo distance ($\sim$ 2.53 $\mathrm \AA$) than
Molybdenum metal ($\sim$ 2.7 $\mathrm \AA$).

We use the first-principles plane-wave pseudopotential method as implemented is
Quantum Espresso package \cite{giannozzi2009quantum} to calculate the
properties of these materials and their heterostructures.  We have used
the recently developed Optimized Norm-Conserving Vanderbilt (ONCV) pseudopotentials
\cite{schlipf2015optimization} and the local density approximation (LDA)
\cite{perdew1981self} for exchange-correlation potential in all our
calculations. The wavefunctions in these calculations is expanded in terms of
plane-waves of energy upto 100 Ry. We chose 4$\times$4$\times$2 k-grid for
sampling the Brillouin zone in case of unit cell with 26 atoms. 
To obtain the equilibrium lattice constants as well as structural parameters we
use DFT to compute the Hellmann-Feynman forces on the atoms and pressure on the
boundaries of the periodic cell. We find that the equilibrium structure when
the forces on atoms are less than $0.01 \mathrm{eV/\AA}$ and the pressure is
less than $0.5 \mathrm{kBar}$. We have used the Berry phase approach
\cite{king1993theory,bernardini1997spontaneous,ederer2005effect,dal1994ab} to
study the spontaneous polarization as well as piezoelectric properties of
A$_2$Mo$_3$O$_8$ compounds in this work (See supplementary materials for
details).For the Berry phase calculation, we find that a 4$\times$4$\times$6
k-grid is sufficient to converge our results.  

The lattice parameters obtained from our calculation is in good agreement with
experimental results \cite{mccarroll1957some} (Table. S1). Both in-plane and
out-of-plane lattice parameters follow a similar trend, MMO$<$ZMO$<$CMO.
Furthermore, the in-plane lattice parameters for ZMO and MMO are very close.
This suggests that an epitaxial heterostructure of ZMO/MMO will have negligible
strain due to lattice mismatch (0.1$\%$). The lattice mismatch between ZMO-CMO
or MMO-CMO while not as small as ZMO-MMO, is still quite low ($\sim $1$\%$)
suggesting low epitaxial strains for these heterostructures as well.
 
In Fig.~\ref{fig1}(c) we show the orbital-resolved density of states (DOS) for
A$_2$Mo$_3$O$_8$ compounds. It shows that both valence and conduction band
edges of these materials are formed as a result of hybridization between
Mo($4d$) and O($2s$,$2p$) states. It has been shown in the literature
\cite{biswas2017optical} that in case of ZMO, by hybridizing Mo($4d$) and
O($2s$,$2p$) states one can construct the bands near valence and conduction
band edge.  Fig.~\ref{fig1}(c) shows that in all these compounds, the A$^{2+}$
cation does not contribute significantly to the states near the Fermi level.
Moreover experimental observations also suggest that the photo-electrochemical
properties of these materials are very similar
\cite{paranthaman1988photoelectrochemical}.  Our results of orbital-resolved
DOS are consistent with experimental observations. The band-gap of
A$_2$Mo$_3$O$_8$ compounds have also been found to be very similar (Table. S1).  

Table.~\ref{tab1} shows the calculated piezoelectric constants and spontaneous
polarizations of A$_2$Mo$_3$O$_8$ compounds. Spontaneous polarizations
of these materials are larger than Group III nitrides
\cite{bernardini1997spontaneous} and close to some of the perovskites such as
BaTiO$_3$ \cite{shieh2009hysteresis} ($0.26 \mathrm{C/m^2}$) and KNbO$_3$
\cite{gunter1977spontaneous} ($0.3-0.4 \mathrm{C/m^2}$). The
piezoelectric constants have an electronic and an ionic contribution.
Electronic or clamped-ion contributions \cite{bernardini1997spontaneous}
(e$_{33}^{(0)}$/e$_{31}^{(0)}$) have been computed by calculating the
piezoelectric response of an applied strain, keeping the atoms fixed at their
equilibrium positions and shown in Table.~\ref{tab1}. The rest of the
piezoelectric response comes from the displacement of atoms in response to
applied strain and also depends on the Born effective charges of those atoms
\cite{bernardini1997spontaneous,dal1994ab}. As we can see from Table.~\ref{tab1} 
the calculated piezoelectric constants of A$_2$Mo$_3$O$_8$ compounds
are consistently smaller when compared with other known piezoelectric materials
\cite{bernardini1997spontaneous,dal1994ab,ederer2005effect}. This is 
primarily due to small ionic contribution, resulting from the cancellation
of piezoelectric response coming from different atoms. Such a cancellation is
absent in wurtzite or perovskite crystal structure as there is only one
structural parameter.  

\begin{table}
\caption{\label{tab1} Calculated values of piezoelectric constants and spontaneous polarization values for A$_2$Mo$_3$O$_8$ compounds}
\begin{tabular}{|c|c|c|c|c|c|}
 \hline
        Material &  e$_{33}$ & e$_{33}^{(0)}$ & e$_{31}$ & e$_{31}^{(0)}$ & $P^{eq}$ (C m$^{-2}$) \\
 \hline
 \hline
 ZMO &  -0.10 & -0.15 & -0.15 & 0.07 & -0.195 \\
 \hline
 MMO &   0.01 & -0.12 & -0.12 & 0.06 & -0.223 \\
 \hline
 CMO &  -0.20 & -0.09 & -0.22 & 0.02 & -0.141 \\
 \hline
\end{tabular}
\end{table}

To compute the properties A$_2$Mo$_3$O$_8$/B$_2$Mo$_3$O$_8$ interface and to
show the formation of 2DEG, we construct a heterostructure consisting of
1$\times$1$\times$3 supercell of A$_2$Mo$_3$O$_8$ and 1$\times$1$\times$3
supercell of B$_2$Mo$_3$O$_8$, stacked along $(0001)$ direction
\cite{betancourt2013polarization}. We fix the in-plane lattice parameter to the
A$_2$Mo$_3$O$_8$ equilibrium value.  We used a 4$\times$4$\times$1 k-grid for
calculation of all the heterostructure properties.  All other computational
details are the same as those for the bulk calculations.  The macroscopic
average electrostatic potential \cite{colombo1991valence} ($\bar{V}(z)$), the
average electric field inside the materials ($\bar{E}(z)=-\frac {\partial
\bar{V}(z)} {\partial z}$) and the average charge ($\bar{\rho}(z)=-\frac
{\epsilon_0\partial^2 \bar{V}(z)} {\partial z^2}$)
\cite{betancourt2013polarization} were computed from the total electrostatic
potential of the supercell.

\begin{figure}[htb!]
\includegraphics[width=8cm] {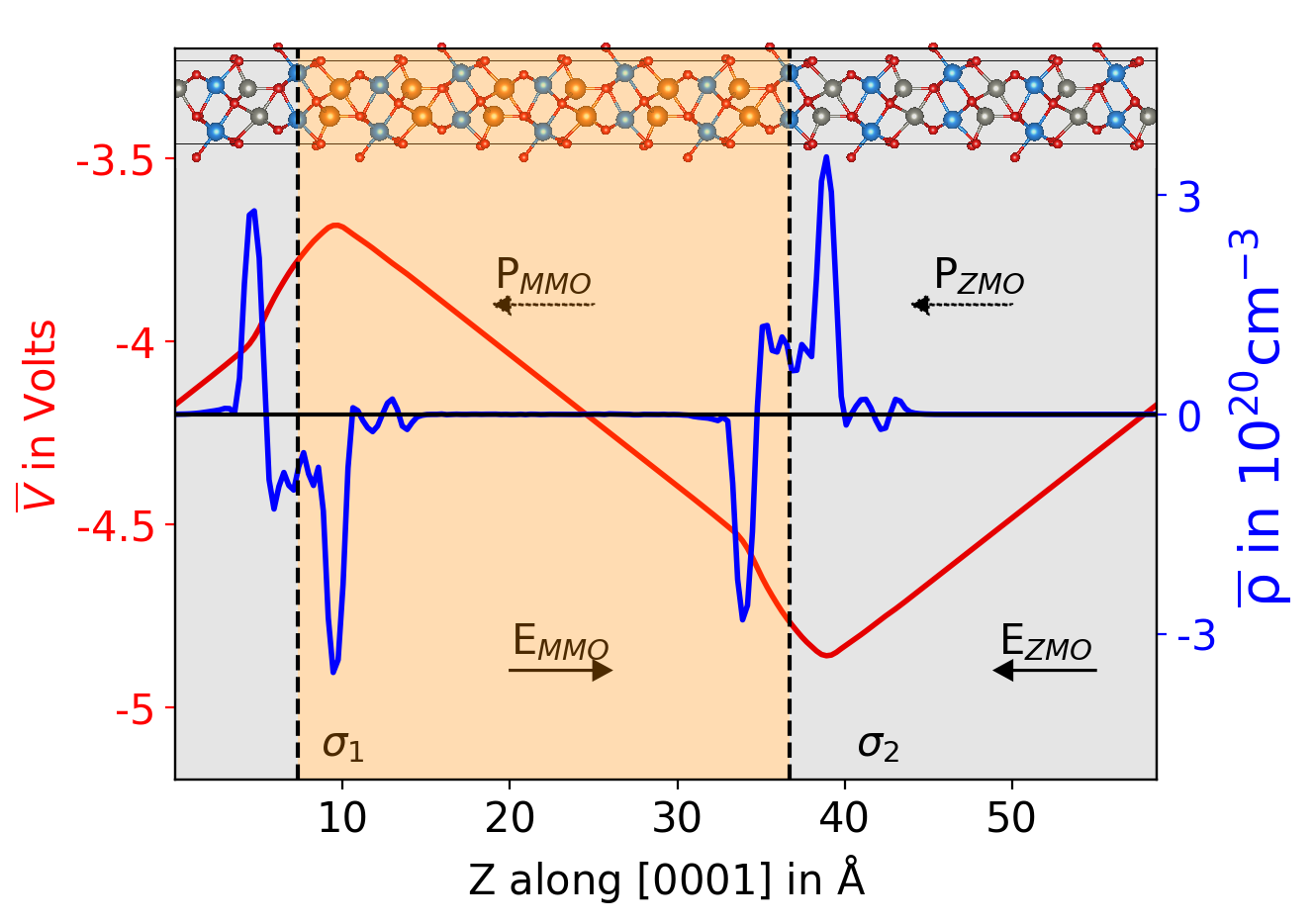}
\caption{Plane-averaged electrostatic potential profile (blue)
$\overline{V}(z)$ and total charge density (red)
$\overline{\rho}(z)$ along the [$0001$] direction in the ZMO/MMO
heterostructure. The zero-field polarization and electric field directions are
indicated by arrows.}
\label{fig2}
\end{figure}

In Fig.~\ref{fig2} we show the calculated macroscopic averaged total charge
density $\overline{\rho}(z)$ and electrostatic potential profile
$\overline{V}(z)$ along the [$0001$] direction obtained directly from ZMO/MMO
heterostructure calculation. The procedure of macroscopic averaging is to wash
out unwanted periodic oscillations \cite{colombo1991valence} coming from
lattice periodicities  of each constituent materials. The macroscopic averaged
electrostatic potential shows the linear behaviour in the bulk of the material,
indicating constant electric field. From the slope of $\overline{V}(z)$ in the
linear region, we calculate the electric field inside the material (Table. S2). 
Moreover, the total charge density $\overline{\rho}(z)$, shows 
charges only close to the interface but not in the bulk regions. 
The surface charge densities at heterostructure interface have been calculated
by integrating $\overline{\rho}(z)$ (Table. S2).

We then proceed to validate our polarization discontinuity hypothesis, by
calculating the electric field and surface charges from electrostatic model.
The charges at the heterostructure interface and the field inside the material
can be computed from polarization discontinuity hypothesis using the
electrostatic boundary conditions \cite{betancourt2013polarization}. Consider a heterostructure
with two materials of length $l_1$ and $l_2$ having dielectric constants
$\epsilon_1$ and $\epsilon_2$ and zero field polarization $P^0_1$ and $P^0_2$,
respectively. Using periodic boundary condition, which ensures
the net potential difference across the supercell is zero, the bound charges at
the interface, $\sigma$, and the field inside the bulk of the materials, $E_1$
and $E_2$, are given as  \cite{betancourt2013polarization},
\begin{equation}
\sigma=-\frac {\Delta P^0} {\bar{\epsilon'}} ; E_1 = -\frac {\Delta P^0} {\bar{\epsilon}} (\frac {l_2} {L}) ; 
E_2 = -\frac {\Delta P^0} {\bar{\epsilon}} (\frac {l_1} {L}) 
\label{ESB}
\end{equation}
where,
\begin{equation}
\Delta P^0 = P^0_1-P^0_2; 
\bar{\epsilon} = \epsilon_1 (\frac {l_2} {L}) + \epsilon_2 (\frac {l_1} {L});
\bar{\epsilon'} = \frac {\bar{\epsilon}} {\epsilon_0}
\end{equation}

We calculate the electronic dielectric constant by applying a small static
homogeneous electric field ($0.3\times10^9 V/m$) along [$0001$] direction
and studying the response of the system using modern theory of polarization
\cite{umari2002ab,souza2002first}(Table. S2). The polarization discontinuity
($\Delta P$) is sum of piezoelectric ($\delta P$) contribution developed in the
MMO layer due to lattice mismatch strain and the spontaneous polarization
($P^{eq}$) difference between ZMO and MMO. As the in-plane lattice parameters
of ZMO and MMO are very close and the e$_{31}$ value of MMO is quite small ($-0.12
\mathrm {C/m^2}$), the piezoelectric effect is negligible. Using Eq.~\ref{ESB},
we find the surface charge density to be $0.403\times10^{13} cm^{-2}$ which
agrees perfectly with the value we find from heterostructure calculation
earlier. Moreover, we find the electric fields inside each slab (E$_{ZMO}$ =
$-0.364\times10^9 V/m$, E$_{MMO}$ = $-0.364\times10^9 V/m$ ) are also in very
good agreement. See supplementary material for these comparisons (Table. S2).  

\begin{figure}[htb!]
\setlength{\unitlength}{0.1\textwidth}
\begin{picture}(4,4)
\put(-0.5,0.1){\includegraphics[width=8cm] {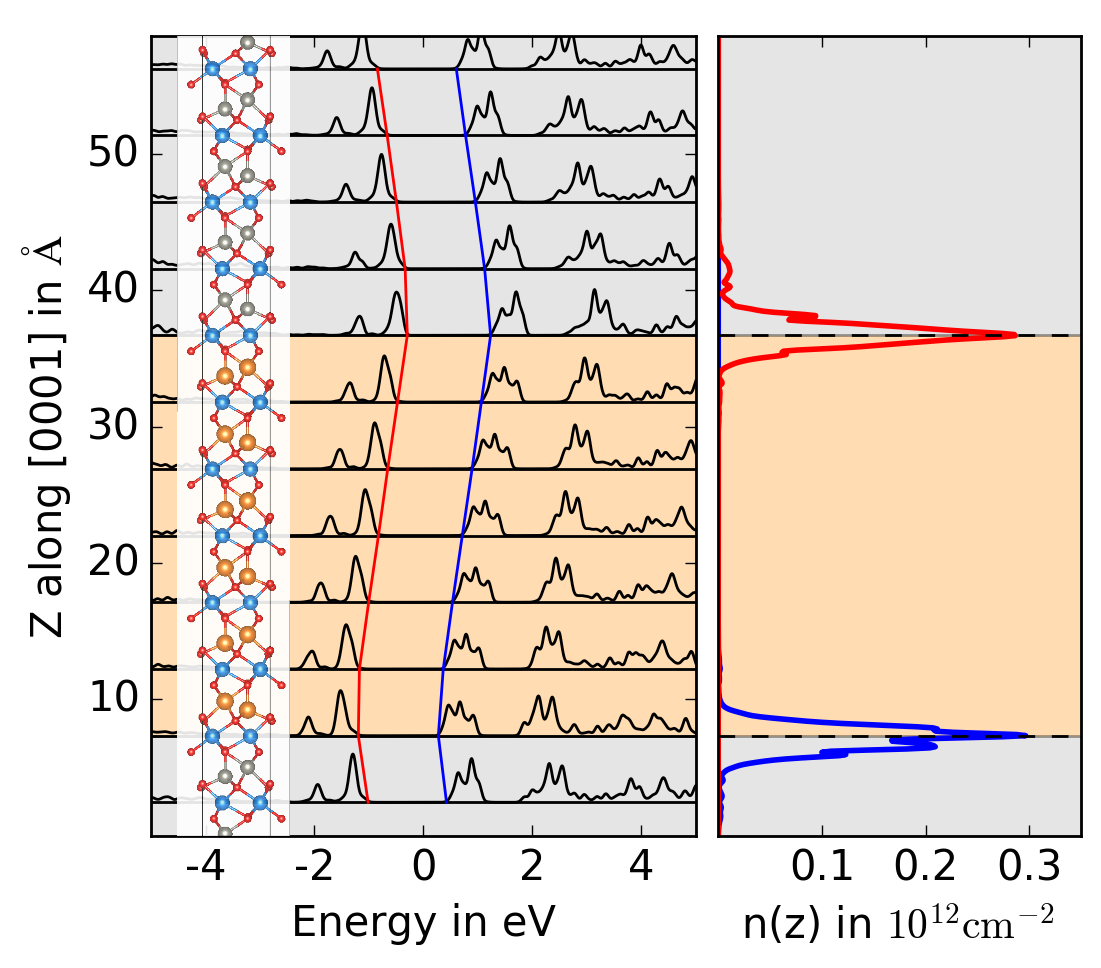}}
\put(1,0){(a)}
\put(3,0){(b)}
\end{picture}
\caption{(a) Calculated layer-resolved Mo(4d) DOS for ZMO/MMO heterostructure.
Red and blue lines are indicating the Valence and Conduction band edge profile
along [$0001$] direction.(b) Charge density distribution of 2DEG (blue) and 2DHG
(red) for ZMO/MMO heterostructure.}
\label{fig3}
\end{figure}

The agreement between results obtained from heterostructure calculation and
from Eq.~\ref{ESB}  which is derived assuming polarization discontinuity
hypothesis proves that the surface charge density in this heterostructure is
indeed due to polarization discontinuity at the interface. The electric field
created by these bound surface charges bring any free carriers created inside
the bulk of the material to the interface making a 2DEG (or 2DHG).
It is important to note that in a real material the formation of 2DEG (or 2DHG)
in some cases may be hindered by formation of defects such as Oxygen vacancies
\cite{zhong2010polarity}. In Fig.~\ref{fig3}(a) we show the layer-resolved Mo(4d)
DOS. We have also performed the same calculation with increasing the thickness
of both ZMO as well MMO layers (Fig. S3). We choose Mo(4d) states because both
the valence and conduction band edge of all A$_2$Mo$_3$O$_8$ compounds are
mostly of Mo(4d) character (Fig.~\ref{fig1}(c)). The layer-resolved Mo (4d) DOS
clearly shows the shift in the valence and conduction bands due to the in-built
electrostatic field inside the heterostructure as we go along [$0001$]
direction. From Fig.~\ref{fig3}(a) (also from layer dependent band structure in 
Fig. S2) it is evident that conduction and valence band edge of the heterostructure
is composed of states localized near one of the two interfaces in the
heterostructure. It should be noted that the linear nature of the potential
inside the bulk region is the consequence of the absence of free carriers in
our calculation.  In real materials free carriers will screen the bound surface
charges at the interface.  As a results one will not see a linear region inside
the material as in Fig.~\ref{fig2}.  Instead the potential will saturate as one
goes away from the interface \cite{betancourt2013polarization} such that the
field far from the interface becomes zero. Nevertheless, free carriers will be
localized at the interface by the electric field of polarization induced bound
charges. In principle, free carriers in real materials can be generated both
spontaneously or as a result of modulation doping. In GaAs/Al(Ga)As
heterostructure, the free carriers are provided by the modulation doping. In
case of GaN/Al(Ga)N heterostructure it has been observed that donor states at
the surface can provide free carriers in the system
\cite{heikman2003polarization}.  The origin of free carriers, while
interesting, is outside the scope of the present study.

To study the localization of the 2DEG (or 2DHG) in the ZMO/MMO heterostructure,
we simulate the addition of free carriers in the system by moving the Fermi
level upwards into the conduction band ($n$-doping) or downwards into the
valence band ($p$-doping).  We move the Fermi level of the system such that we
add (remove) 0.003 $e$ to (from) the heterostructure supercell. This
corresponds to a surface charge density of $\sim$ $1\times10^{12} cm^{-2}$. In
Fig.~\ref{fig3}(b) we plot $\sum_{n,k} |\psi_{n,k}(z)|^{2}$ where $\psi_{n,k}$s
are the occupied (emptied) states as a result of doping. Blue and red lines
have been used to show the electron density in case of $n$-doping and hole
density is case of $p$-doping respectively.  As one can see from
Fig.~\ref{fig3}(b), both the 2DEG as well as 2DHG is well localized to within
$<10\mathrm{\AA}$ from the interface. The strong localization at the interface
is a consequence of the fact that the excess carriers due to doping occupy
interfacial Mo(4d) states which are localized in the [0001] direction.

\begin{table}
\caption{\label{tab2} Surface charge density for different A$_2$Mo$_3$O$_8$/B$_2$Mo$_3$O$_8$ interface
Absolute values of $\Delta P$ has been reported.}
\begin{tabular}{|c|c|c|}
 \hline
 Interface &  $\Delta P$ in $C/m^2$ & $\sigma$ in $10^{13} cm^{-2}$ \\
 \hline
 \hline
  ZMO/MMO & 0.028 & 0.403 \\
 \hline
  ZMO/CMO & 0.053 & 0.647 \\
 \hline
  MMO/CMO & 0.082 & 1.088 \\
 \hline
\end{tabular}
\end{table}

In Table.~\ref{tab2} we list the surface charge densities for different
possible A$_2$Mo$_3$O$_8$/B$_2$Mo$_3$O$_8$ interface. The values are calculated
using Eq.~\ref{ESB}, as we expect polarization discontinuity driven 2DEG (or
2DHG) to form in case of other heterostructure like ZMO/CMO and MMO/CMO as
well. We can see the value of surface charge density is highest for MMO/CMO
interface due to largest polarization discontinuity. These values are
comparable to other well known heterostructure systems that has been found to
host 2DEG such as GaN/Al(Ga)N or ZnO/Zn(Mg)O
\cite{heikman2003polarization,tampo2008polarization}, in which the
surface charge densities were found to be of the order of $10^{13} cm^{-2}$.
This suggests that the A$_2$Mo$_3$O$_8$/B$_2$Mo$_3$O$_8$ interface can be a
strong candidate for hosting 2DEG/2DHG in an all oxide system.

In this work we explore the possibility of 2DEG formation in a novel
heterostructure system.  The materials forming these heterostructure has the
chemical formula A$_2$Mo$_3$O$_8$ where A can be Zn, Mg or Cd. All these
materials have been synthesized before and found to be very stable. We
calculate the piezoelectric properties of these materials by applying an
external strain and studying the response of the system. We also compute the
value of spontaneous polarization of these materials using Berry phase method.
We then proceed with DFT calculation of slab based heterostructure system
consists of ZMO and MMO. We show that there are localized surface charges at the
interface and electrostatic field inside the material. We show that these bound
charges are due to polarization discontinuity at the interface. We show
excellent agreement of surface charge density and electric field values between
heterostructure calculation and polarization discontinuity model. We then
simulate doping of the system by moving the fermi level of the system and show
that the additional charges are localized within $<10\mathrm{\AA}$ from the
interface. We also report the values of surface charge densities for other
possible heterostructure system of these materials such as ZMO/CMO or MMO/CMO.
We show that the surface charge densities in these systems are comparable to
other well known heterostructure system that forms 2DEG, such as AlN/Al(Ga)N or
ZnO/Zn(Mg)O.  

\section{Acknowledgment}
The authors thank Prof. Rajeev Ranjan, Prof. Srimanta Middey, Prof. Sumilan
Banarjee, Prof. S. Raghavan and Prof. S. Avasthi for useful discussions. This
work is supported under the US-India Partnership to Advance Clean
Energy-Research (PACE-R) for the Solar Energy Research Institute for India and
the United States (SERIIUS), funded jointly by the U.S. Department of Energy
(Office of Science, Office of Basic Energy Sciences, and Energy Efficiency and
Renewable Energy, Solar Energy Technology Program, under Subcontract
DE-AC36-08GO28308 to the National Renewable Energy Laboratory, Golden,
Colorado) and the Government of India, through the Department of Science and
Technology under Subcontract IUSSTF/JCERDC-SERIIUS/2012 dated 22nd Nov. 2012.
We thank Super Computer Research and Education Centre (SERC) at IISc for the
computational facilities.

\bibliography{piezo} 

\end{document}

\affiliation{Center for Condensed Matter Theory, Department of Physics, Indian Institute of Science, Bangalore, 560012}

\author{Tathagata Biswas}
\affiliation{Center for Condensed Matter Theory, Department of Physics, Indian Institute of Science, Bangalore, 560012}
\author{Manish Jain}
\affiliation{Center for Condensed Matter Theory, Department of Physics, Indian Institute of Science, Bangalore, 560012}
\email{mjain@physics.iisc.ernet.in}

\title{Supporting Information for the Manuscript Titled: ``Polarization
discontinuity driven two dimensional electron gas at
A$_2$Mo$_3$O$_8$/B$_2$Mo$_3$O$_8$ (A, B : Zn,Mg,Cd) interfaces''} 
\maketitle

\section{Computational details}
The Berry phase approach allows one to calculate the difference in polarization between two states of a
system, provided they can be connected through an adiabatic transformation
which keeps the system insulating throughout the process \cite{king1993theory}.
The difference in electronic polarization $\Delta P_e$ between two systems can then
be calculated from the geometric quantum phase as
\cite{king1993theory,bernardini1997spontaneous}:
\begin{eqnarray}
\Delta P_e &=& P_e(\lambda_2) - P_e(\lambda_1) \\
P_e(\lambda) &=& - \frac {2 e} {(2 \pi )^3 } \int_{BZ} d{\bf k} \frac {\partial} {\partial {\bf k'} } \phi ^{(\lambda)}({\bf k,k'}) |_{{\bf k=k'}}  
\end{eqnarray}
where the domain of integration is the reciprocal-lattice unit cell and
$\lambda$ is a parameter which is changed continuously to transform a
structure labeled by $\lambda_1$ adiabatically to one that is labeled by
$\lambda_2$. The geometric phase, $\phi ^{(\lambda)}$ can be computed
from the occupied  Bloch states of the crystal ($u^{(\lambda)}_n({\bf k})$)
using,
\begin{equation}
\phi ^{(\lambda)}({\bf k,k'}) =  Im ( ln [det \langle u^{(\lambda)}_m({\bf k}) | u^{(\lambda)}_n({\bf k'})])
\end{equation}
It is important to note that the geometric quantum phase is only defined modulo
2$\pi$.  As a result the polarization is also only defined modulo $\frac {e{\bf
R}} {\Omega}$, where $\bf R$ is the real-space lattice vector in the direction
of polarization.

The total macroscopic polarization ($\bf P$) of a solid is the sum of
spontaneous polarization ($\bf P ^{eq}$) of its equilibrium structure and
piezoelectric polarization ($\delta \bf P$) induced as a result of any applied
strain ($\epsilon$). Within the Berry phase approach, the spontaneous polarization
is calculated with respect to a structure that has zero polarization and is an insulator. In case of
materials with wurtzite crystal structure, the zinc blend structure of the same material is
a natural choice. The difference in polarization
obtained using this reference structure has been found to be in very good agreement
with the experimental values \cite{dal1994ab,bernardini1997spontaneous}.
However, for the materials that we are studying, no known inversion symmetric
structure exists. As a result, we use a hypothetical crystal structure, obtained by
moving the atoms of the original structure to restore inversion symmetry as a reference.
We note that we are {\em assuming} that one can relate the two
structures via a gap preserving adiabatic transformation just like in case of
previous studies \cite{dal1994ab,bernardini1997spontaneous}.

The piezoelectric polarization ($\delta {\bf P}$) within linear response (using
Voigt notation) can be written \cite{fast1995elastic} as,
\begin{equation} 
\delta {\bf P}_i = e_{ij} \epsilon_{j} 
\end{equation}
where is $e_{ij}$ is the piezoelectric tensor. As the A$_2$Mo$_3$O$_8$
materials have a hexagonal crystal structure and we are only interested in
polarization along [$0001$] direction (${\bf P}^{eq} =P^{eq} {\bf \hat{z}}$),
there are only two independent components of piezoelectric tensor, $e_{33}$ and
$e_{31}$ \cite{bernardini1997spontaneous}. We are not considering any shear
strain here, so $e_{51}=0$. Piezoelectric polarization in this case can be
written as,
\begin{equation}
\delta P_3 = e_{33} \epsilon_3 + e_{31} (\epsilon_1 + \epsilon_2) 
\end{equation}
where $\epsilon_3= (c-c_0)/c$ is the strain along $c$-axis and $\epsilon_1=
\epsilon_2= (a-a_0)/a$ is the in-plane strain.  The equilibrium lattice
parameters of the systems are $a_0$ and $c_0$ along the in-plane and $c$-axis
respectively.  Using a Taylor expansion one can also write the $\delta P_3$ as,
\begin{equation}
\delta P_3 = \frac {\partial P_3} {\partial a} \bigg{|}_{a=a_0}  (a-a_0) + \frac {\partial P_3} {\partial c} \bigg{|}_{c=c_0} (c-c_0)
\end{equation}
It is important to note that we do not consider internal structural parameters
as independent variables in the above equation. This is because in our
calculations, we calculate the polarization of the relaxed structure with the
constrained lattice parameters \cite{ederer2005effect,dal1994ab}. Using the
above equations, one can compute the piezoelectric constants as,
\begin{equation}
e_{33} = c_0 \frac {\partial P_3} {\partial c} \bigg{|}_{c=c_0}  ;  \; \; e_{31} = \frac {a_0} {2} \frac {\partial P_3} {\partial a} \bigg{|}_{a=a_0} 
\end{equation}
In practice, we apply a small strain ($\pm 1 \%$) along $c$-axis (to calculate
$e_{33}$) or in $xy$-plane (to calculate $e_{31}$) and calculate the
polarization of the relaxed structure at the strained lattice parameters.
In the small strain limit, the polarization is linear with respect to the
strain. The piezoelectric constants can thus be simply calculated from the
slope of the polarization vs strain curve.

\section{Crystal structure of A$_2$Mo$_3$O$_8$ compounds and the inversion symmetric 
structure used for calculating spontaneous polarization}

\begin{figure*}[h!]
\setlength{\unitlength}{0.1\textwidth}
\begin{picture}(6,4.8)
\put(1,2.7){\includegraphics[height=4cm] {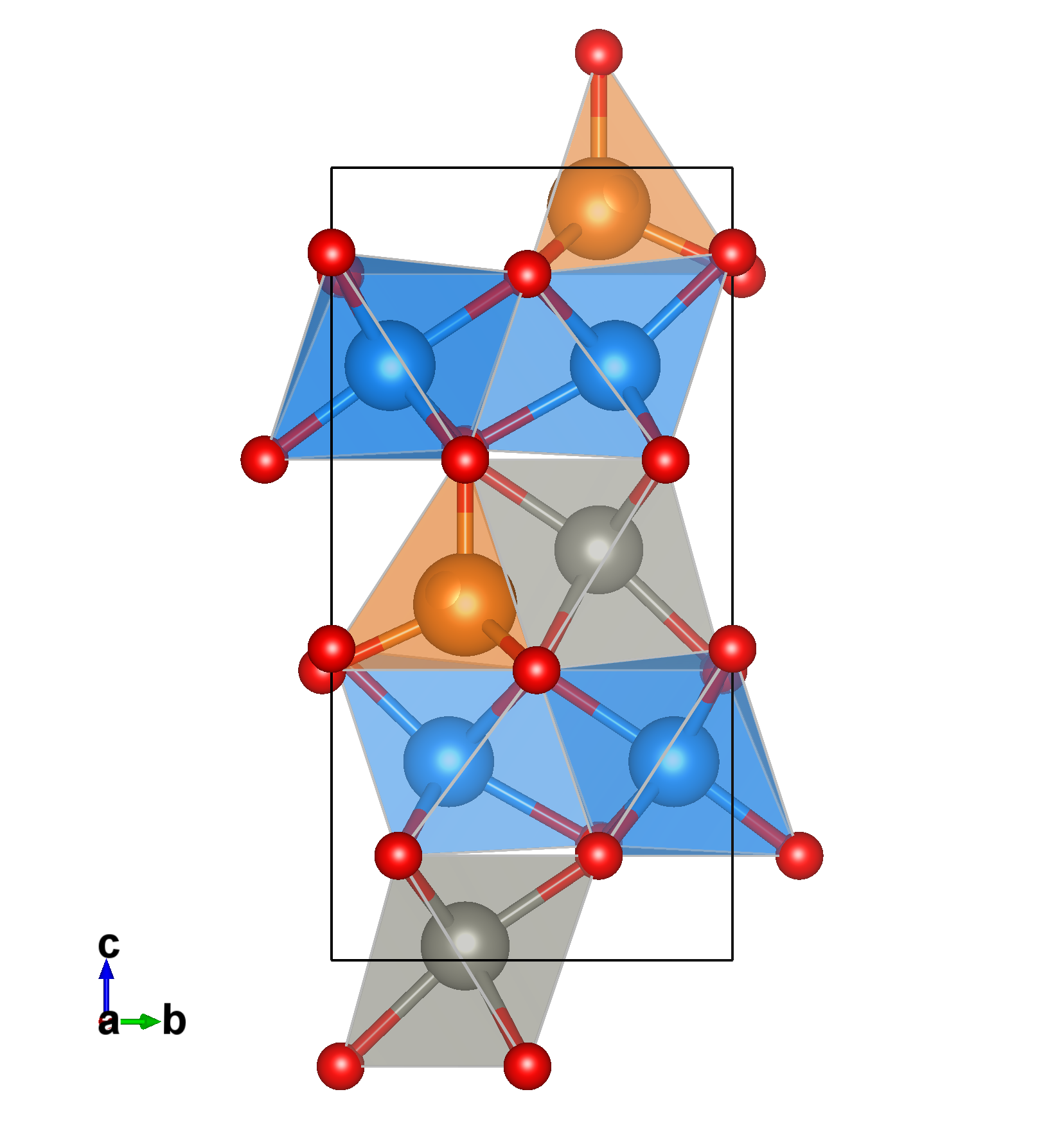}}
\put(3,2.7){\includegraphics[height=4cm] {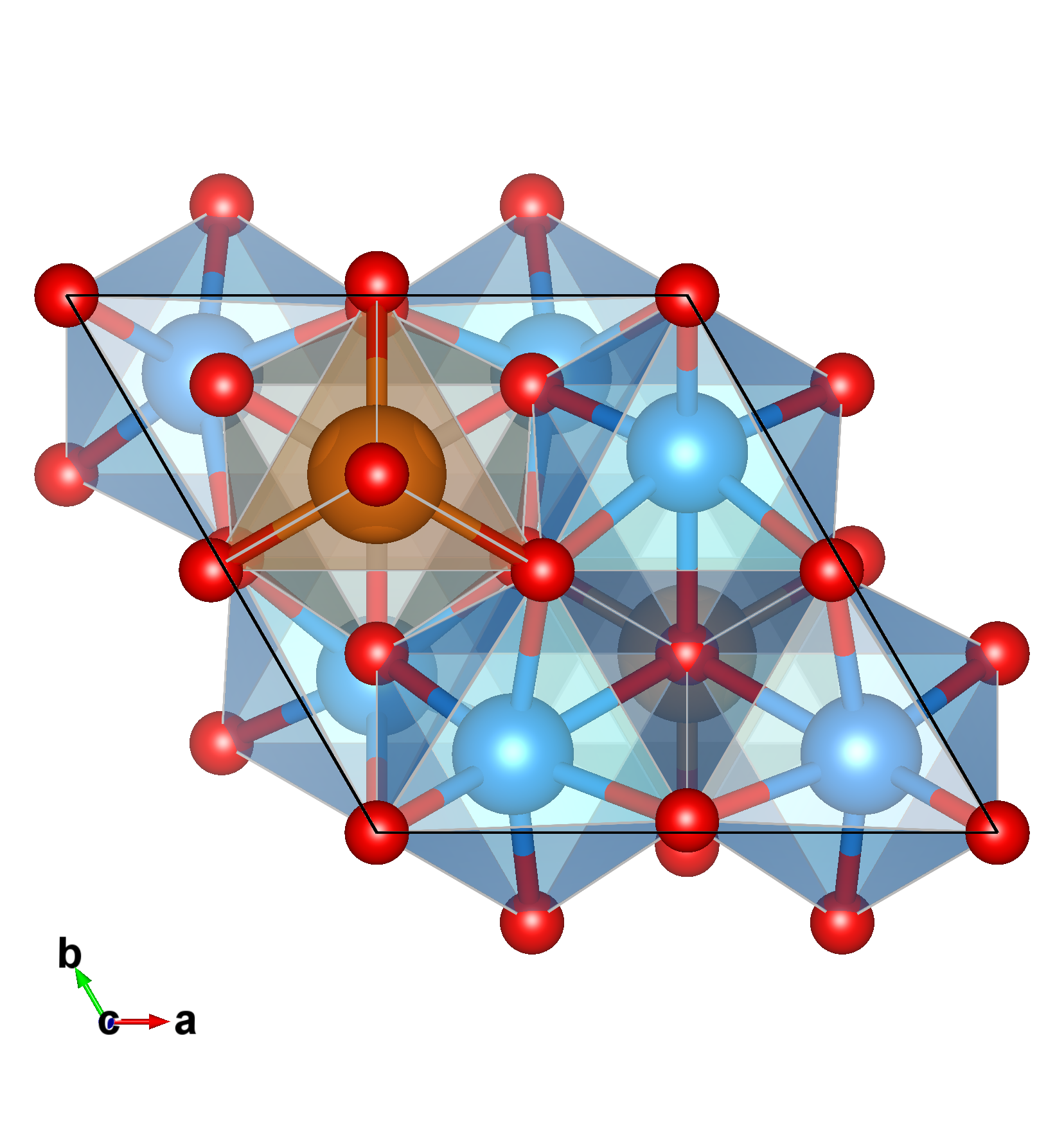}}
\put(1,0.2){\includegraphics[height=4cm] {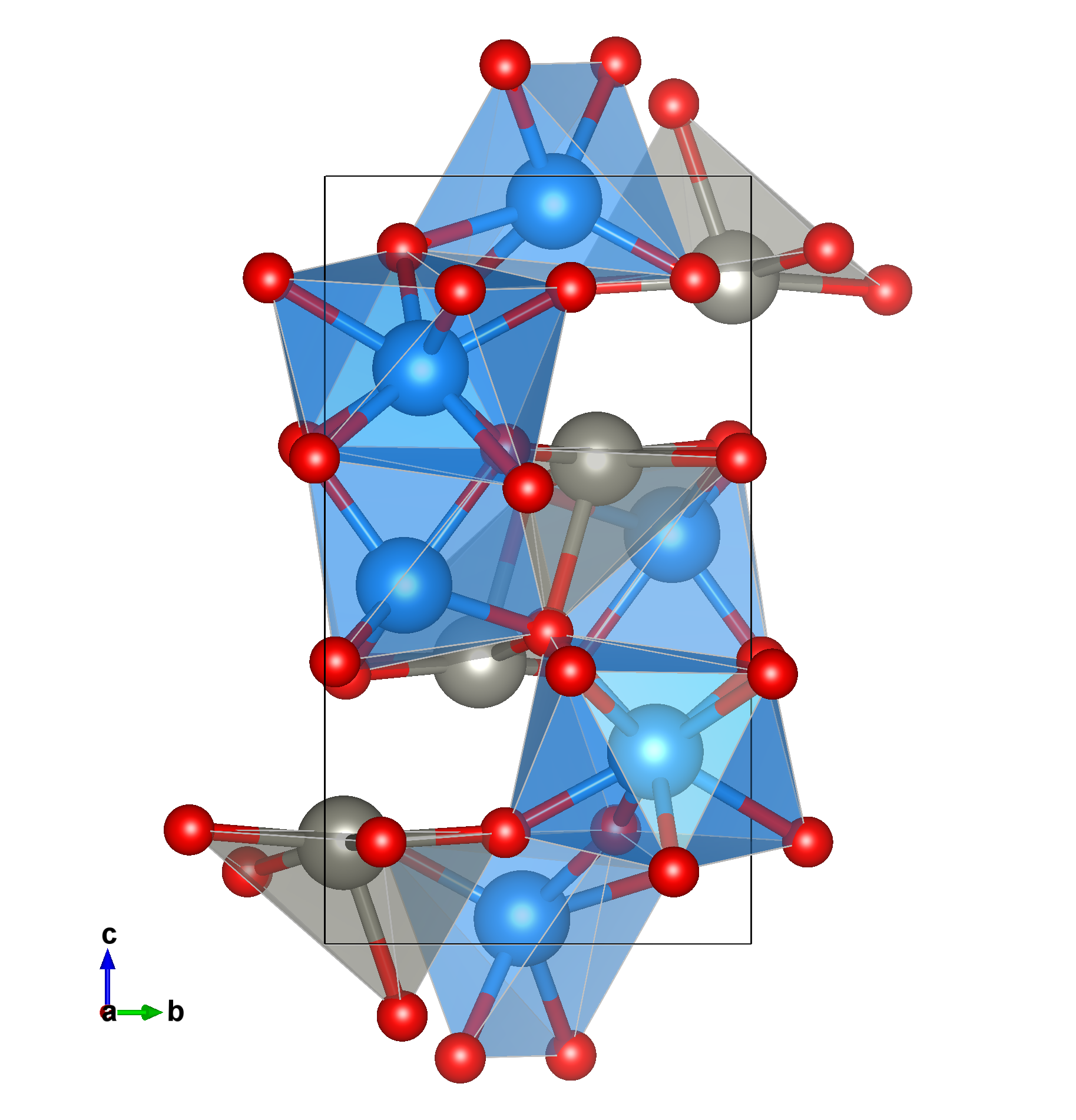}}
\put(3,0.2){\includegraphics[height=4cm] {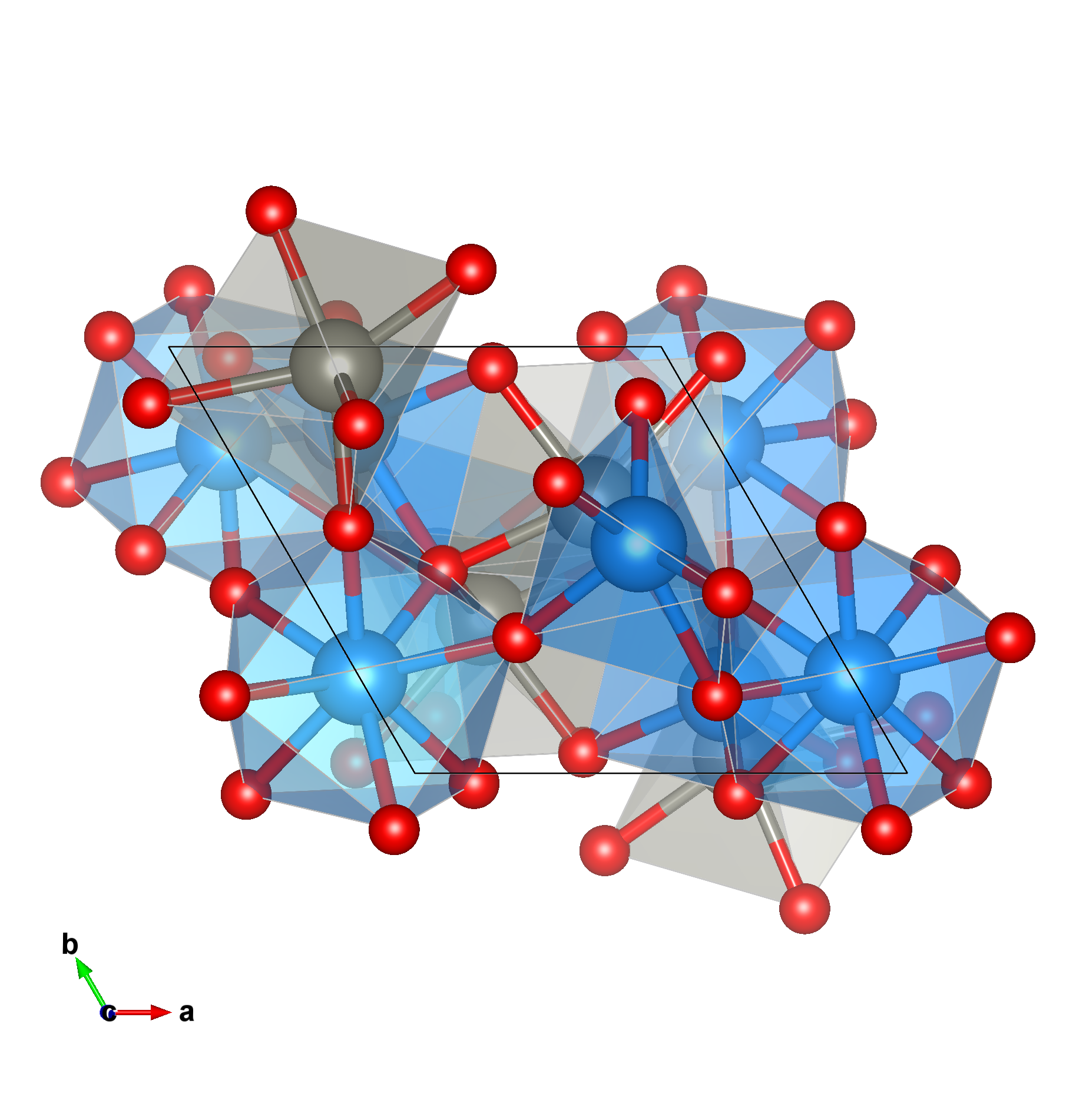}}
\put(2,2.5){(a)}
\put(4,2.5){(b)}
\put(2,0){(c)}
\put(4,0){(d)}
\end{picture}
\caption{ Crystal structure of A$_2$Mo$_3$O$_8$ compounds from (a)
[$2\bar{1}\bar{1}0$] (b) [$0001$] direction.  Yellow, grey, blue and red
spheres represents A$_{tetra}$, A$_{octa}$, Mo and O atoms respectively. 
The inversion symmetric crystal structure used as reference to calculate
spontaneous polarization from (c) [$2\bar{1}\bar{1}0$] (d) [$0001$] direction.} 
\label{fig1}
\end{figure*}

\begin{table*}[h!]
\caption{\label{tab1} Lattice parameters, Wycoff positions and band gaps for
A$_2$Mo$_3$O$_8$ compounds. Both calculated values and experimental results
\cite{mccarroll1957some} are shown for lattice parameters.}

\begin{tabular}{|m{3cm}|m{3.5cm}|m{3.5cm}|m{3.5cm}|}
\hline
parameters & ZMO & MMO & CMO\\
\hline
\hline
$a^{Theory}$ in \AA & 5.707 & 5.701 & 5.764 \\
\hline
$a^{Expt.}$ in \AA & 5.775 & 5.761 & 5.835 \\
\hline
$c^{Theory}$ in \AA & 9.770 & 9.802 & 10.742 \\ 
\hline
$c^{Expt.}$ in \AA & 9.915 & 9.893 & 10.815 \\ 
\hline
$A^1(2b)$ & (1/3, 2/3, 0.5179) & (1/3, 2/3, 0.5122) & (1/3, 2/3, 0.5148) \\
\hline
$A^2(2b)$ & (1/3, 2/3, 0.9489) & (1/3, 2/3, 0.9479) & (1/3, 2/3, 0.9616) \\
\hline
$Mo(6c)$ & (0.1461, 0.8539, 0.2505) & (0.1463, 0.8537, 0.2507) & (0.1458, 0.8542, 0.2514) \\
\hline
$O^1(2a)$ & (0, 0, 0.8929) & (0, 0, 0.8933) & (0, 0, 0.8821) \\
\hline
$O^2(2b)$ & (1/3, 2/3, 0.1448) & (1/3, 2/3, 0.1459) & (1/3, 2/3, 0.1566) \\
\hline
$O^3(6c)$ & (0.4886, 0.5114, 0.3660) & (0.4873, 0.5127, 0.3669) & (0.4867, 0.5133, 0.3522) \\
\hline
$O^4(6c)$ & (0.1667, 0.8333, 0.6318) & (0.1675, 0.8325, 0.6324) & (0.1613, 0.8387, 0.6411) \\
\hline
Band gap in eV & 1.65 & 1.74 & 1.65\\
\hline
\end{tabular}
\end{table*}

\section{Additional results from ZMO/MMO heterostructure calculations}

\begin{figure}[H]
\centering
\setlength{\unitlength}{0.1\textwidth}
\begin{picture}(6,3)
\put(0,0){\includegraphics[height=6cm] {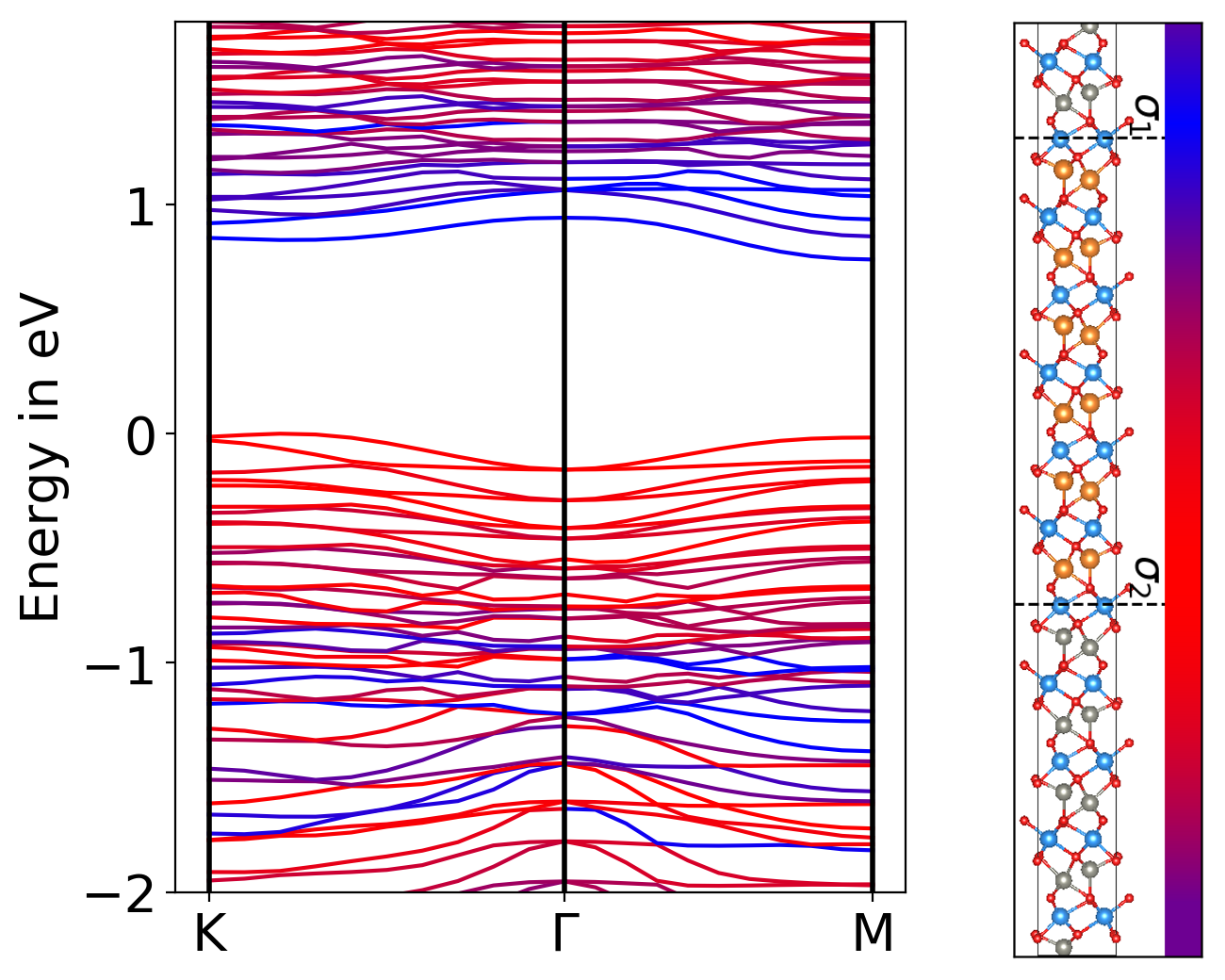}}
\end{picture}
\caption{Layer resolved bandstucture, with the color used to denote which
Mo($4d$) states are involved. The intensity of the blue (red) color decreases
as the distance from $\sigma_1$ ($\sigma_2$) surface increases. To plot the
band structure we chose only those directions in the Brillouin zone which
corresponds to wavevectors parallel to the interface (K[1/3,1/3,0]--$\Gamma$[0,0,0]--M[1/2,0,0])}
\label{fig2}
\end{figure}

\begin{figure}[H]
\centering
\setlength{\unitlength}{0.1\textwidth}
\begin{picture}(6,3)
\put(0,0.1){\includegraphics[height=6cm] {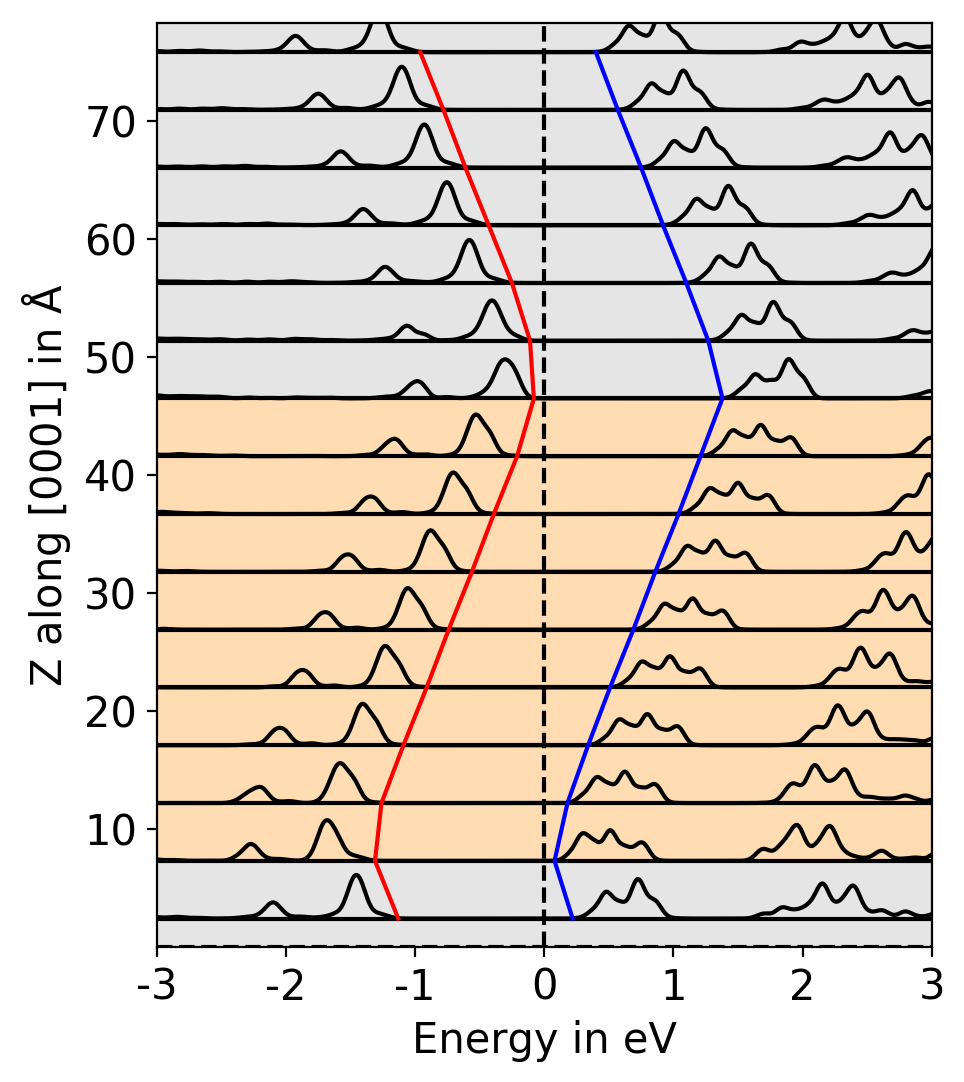}}
\put(3,0.1){\includegraphics[height=6cm] {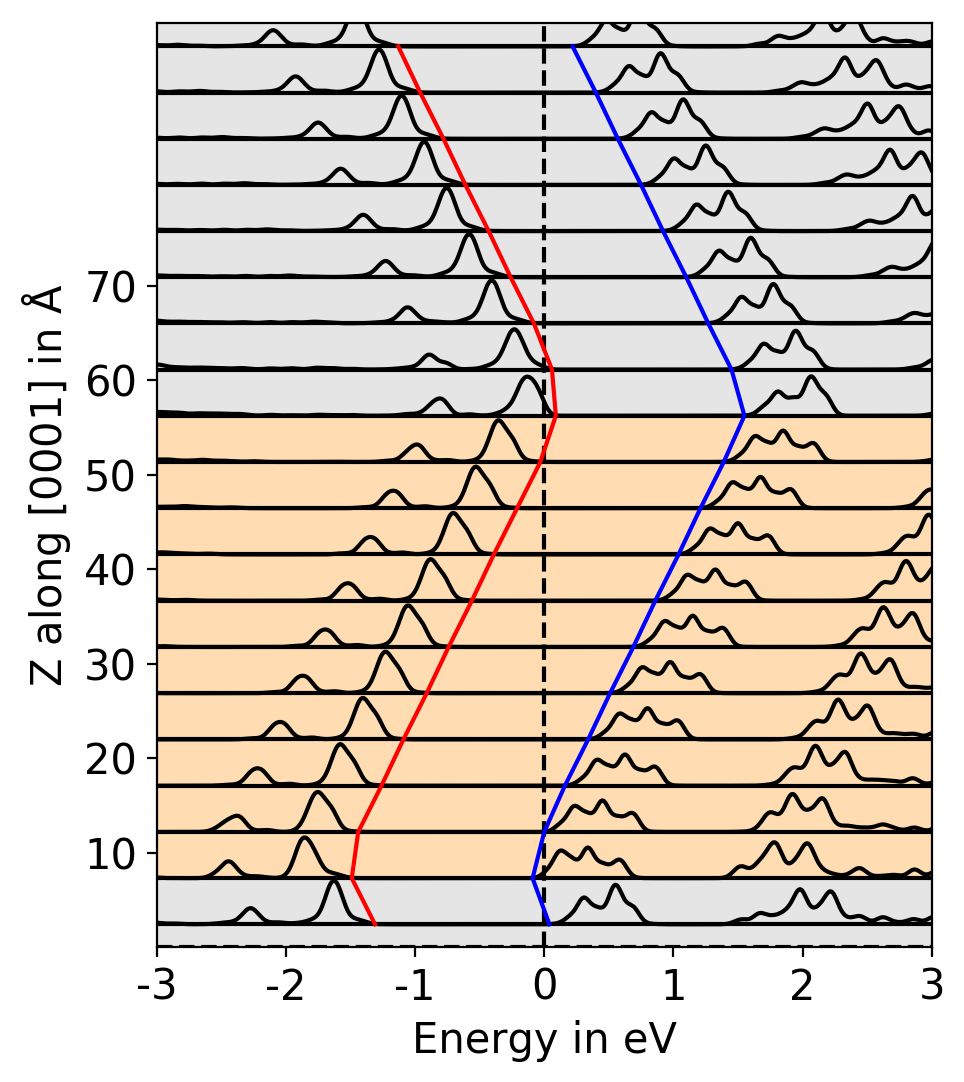}}
\put(0,0){(a)}
\put(3,0){(b)}
\end{picture}
\caption{(a) Calculated layer-resolved Mo(4d) DOS for ZMO/MMO heterostructure
containing 1$\times$1$\times$4 supercell of both ZMO and MMO. (b) The same calculated
using a heterostructure of 1$\times$1$\times$5 supercell of each material.  Red
and blue lines are indicating the Valence and Conduction band edge profile
along [$0001$] direction.}
\label{fig3}
\end{figure}

\section{Results obtained from heterostructure calculation directly
and electrostatic model assuming polarization discontinuity hypotheosis}

\begin{table}[H]
\centering
\caption{\label{tab2} Comparison between results from DFT heterostructure
calculation and electrostatic model for ZMO-MMO interface}
\begin{tabular}{|m{4cm}| m{2cm}| m{2cm}|}
 \hline
 Quantity & DFT  & Model \\
 \hline
 \hline
 $\Delta$P in C m$^{-2} $ & 0.028 & -- \\
 \hline
 $\epsilon_{\rm{ZMO}}$  & 4.894 & -- \\
 \hline
 $\epsilon_{\rm{MMO}}$  & 3.948 & -- \\
 \hline
 $\sigma_{1}$ in 10$^{13}$ cm$^{-2}$ & 0.402 & 0.403 \\
 \hline
 $\sigma_{2}$ in 10$^{13}$ cm$^{-2}$ & -0.401 & -0.403 \\
 \hline
 E$_{\rm{ZMO}}$ in  10$^{9}$ V m$^{-1}$ & -0.358 &  -0.364 \\
 \hline
 E$_{\rm{MMO}}$ in 10$^{9}$ V m$^{-1}$ & 0.359 &  0.363 \\
 \hline
\end{tabular}
\end{table}

\bibliography{supp}